On the role of Loewner entropy in statistical mechanics of 2D Ising system

Yusuke Shibasaki[1,2, *]

[1]*Institute of Natural Sciences, College of Humanities and Sciences, Nihon University, Setagaya, Tokyo 156-8550, Japan*

[2]*Institut des Hautes Études Scientifiques, 35 route de Chartres, Bures Sur Yvette 91440, France*

*yshibasaki98@gmail.com

Abstract

The fundamental properties of 2-dimensional (2D) Ising system were formulated using the Loewner theory. We focus on the role of the complexity measure of the 2D geometry, referred to as the Loewner entropy, to derive the statistical-mechanical relations of the 2D Ising system by analyzing the structure of the interface (i.e., the phase separation line). For the mixing property of the discrete Loewner evolution, we assume that the Loewner driving force $\eta_s(n)$ obtained from the interface has a stationary property, where the autocorrelation function $\langle \eta_s(0)\eta_s(n) \rangle$ converges in the long-time limit. Using this fact, we reconstruct the continuous Loewner evolution driven by the diffusion process whose increments correspond to the sequence of $\eta_s(n)$, and the fractal dimension of the generated curve was derived. We show that these formulations lead to a novel expression of the Hamiltonian, grand canonical ensemble of the system, which also are applicable for the non-equilibrium state of the system. In addition, the relations on the central limit theorem (CLT) governing the local fluctuation of the interface, the non-equilibrium free energy, and fluctuation dissipation relation (FDR) were derived using the Loewner theory. The present results suggest a possible form of the complexity-based theory of the 2D statistical mechanical systems that is applicable for the non-equilibrium states.

# I. INTRODUCTION

The recent attempts on connecting dynamical systems theory and the (chordal) Loewner theory have revealed that the quantification method of the complexity of the 2D geometry of the physical systems in a manner that affords encoding the curve-like 2D morphology on the upper half-plane $\mathbb{H}$ into real-valued 1D time series called the Loewner driving function [1-3]. The Loewner differential equation (LDE) [4, 5] provides a theoretical scheme for such information processing, involving the unique transformation whose essential properties are inherited from that of the conformal mappings [6]. One of the most successful applications of LDE to the problems in physics might be the description of the interface (i.e., the phase separation line) of the 2D Ising system [7, 8]. The discovery of the stochastic Loewner evolution (SLE) in 2000 suggested that the conformally invariant random curves in 2D statistical mechanics models are described by SLE [9], which has a form of LDE driven by the Wiener process $B_t$ parametrized by diffusion parameter $\kappa$. For the 2D Ising system, the interface at the critical temperature $T_c$ has been shown to be described by SLE with $\kappa = 3$ [10], where the scale invariance of the geometry also is most valid. The mathematically rigorous results on the SLE description of the Ising interface are limited to that at the critical temperature $T_c$; however, some studies were performed for analyzing the off-critical Ising interface. i.e., that below $T_c$, where the conformal (and scale) invariance does not hold validly. In this context of the SLE theory, the connection between the statistical mechanical properties of 2D Ising system and dynamical variables in the Loewner theory is not completely clarified. For example, despite the Hamiltonian of the 2D Ising system is closely related to its geometry [7], the link between the quantity in the Loewner theory and Ising interface remains to be an issue. The recent studies of the author numerically suggested that the exponential function-type relation between the Hamiltonian of the system and the entropy of the Loewner driving force, which we referred to as the Loewner entropy [2]. This result is obtained for the system at and below $T_c$, and indicates a possible theoretical formulation of the statistical mechanics of the 2D Ising system using the Loewner theory.

Let us consider the above problem in terms of the complexity. The previous studies of the authors suggested that the Loewner theory aids to work for the quantification of the complexity of the 2D geometry of the Ising system from that of the Loewner driving force, which is a certain type of dynamical system in one-dimension [1, 2]. In general, the complexity of the 2D geometry is quantified by the fractal dimension $d_f$, and this might be a simplest method for measuring the Ising interface. The efficacy of the Loewner theory appears in the sense that it shows the dynamics of the Loewner driving force determine the fractal dimension $d_f$ [3]. In Ref. [3], it has been shown that the fractal dimension $d_f$ of the curve is determined by the autocorrelation function of the Loewner driving force, if we regard it as a dynamical system. It means that, using the Loewner theory, the complexity of the 2D geometry is discussed in terms of analytical calculus that is used in the statistical mechanics. In this manner, the problems in 2D geometry are translated into those in the one-dimensional dynamical system.

In this study, we observe that the formulation of statistical mechanics of 2D Ising system using the Loewner theory. In the formulation, we focus on the Loewner entropy, which works as the complexity measure for the 2D interfaces, based on the assumption that the geometry of the 2D Ising model determines the configuration probability. It should be noted that due to this assumption and mixing property of the Loewner driving force, the formulation using the Loewner entropy also is applicable for the non-equilibrium state of the Ising system.

## II. STATISTICAL MECHANICS OF 2D ISING SYSTEM

Let us consider the 2D Ising system with the lattice $\Omega$ comprised of $N \times N$ segments with Dobrushin boundary condition whose Hamiltonian $H_N$ is given by [7]:

$$H_N = -\frac{1}{2}\sum_{\langle i,j \rangle} \sigma_i \sigma_j - \frac{1}{2}\sum_{i \in \partial^+\Omega} \sigma_i + \frac{1}{2}\sum_{i \in \partial^-\Omega} \sigma_i. \tag{1}$$

Here, $\sigma_i$ denotes the spin $\sigma_i = \{+1, -1\}$ and the sum is taken over all of the pairs of the nearest neighbor included in $\Omega$. The left half of the boundary of $\Omega$ denoted as $\partial^+\Omega$ is fixed as $+1$, and the right half boundary $\partial^-\Omega$ is fixed as $-1$. The Hamiltonian is expressed in terms of the length of the phase separation lines as [2, 7]:

$$H_N = C_N + \text{len}(\gamma_{[0,s]}) + \sum_{k=1}^{n} \text{len}(\lambda_k), \tag{2}$$

where, $C_N$ is a $N$ dependent constant, and $\text{len}(\gamma_{[0,s]})$ denotes the length of the interface passing through the middle point of the bottom base ($\gamma_0 = 0$) and that of the top base ($\gamma_s$) of the lattice $\Omega$. Similarly, $\text{len}(\lambda_k)$ denotes the length of the closed contour line. For the latter discussions, we denote the lattice segment size as $\varepsilon$. The above relation indicates that the complexity of the geometry of 2D Ising model determines the energy of the system. In this scheme, the grand canonical ensemble also is given by the length of the phase separation lines as:

$$P(\gamma_{[0,s]}, \lambda_1, \lambda_2, \ldots, \lambda_n) = \frac{\exp\left[-\beta\left(\text{len}(\gamma_{[0,s]}) + \sum_{k=1}^{n} \text{len}(\lambda_k)\right)\right]}{Z(\Omega, \beta)}. \tag{3}$$

Here, $\beta$ is the inverse temperature and the partition function $Z(\Omega, \beta)$ is expressed as:

$$Z(\Omega, \beta) = \sum_{\gamma_{[0,s]}, \lambda_1, \lambda_2, \ldots, \lambda_n} \exp\left[-\beta\left(\text{len}(\gamma_{[0,s]}) + \sum_{k=1}^{n} \text{len}(\lambda_k)\right)\right]. \tag{4}$$

The above formulation is possible on the assumption that the configuration of the phase separation line and spin configuration have a one-to-one correspondence relation [7]. In the following section, we reformulate the above relations in a manner that measuring the complexity of the 2D interface $\gamma_{[0,s]}$ using the Loewner theory.

### III. FORMULATION USING LOEWNER THEORY

#### A. Loewner entropy from a dynamical system pint of view

In this section, we shall reformulate the statistical-mechanical properties of the 2D Ising system using the Loewner theory. To apply the chordal Loewner evolution to the above-described model, we regard the interface $\gamma_{[0,s]}$ as the points on the upper half-plane $\mathbb{H}$, which is described as $\gamma_{[0,s]} = \{z_0 (= 0), z_1, \ldots, z_n, \ldots, z_L\}$. The continuous version of chordal Loewner evolution is expressed as [1-3, 5, 6]:

$$\frac{\partial g_s(z)}{\partial s} = \frac{2}{g_s(z) - U_s}, \quad g_0(z) = z \in \mathbb{H}. \tag{5}$$

Here, $g_s(z)$ is the map from $\mathbb{H} \setminus \gamma_{[0,s]}$ to $\mathbb{H}$, and $U_s$ is the real-valued driving function, which is one-dimensional. Assuming that the Ising interface $\gamma_{[0,s]}$ is described by Eq. (5), we calculate the Loewner driving function $U_s$. Using the discrete version of the Loewner evolution [11,12], the sequences of the increments of the driving process $\{\Delta U_{s_n}\}$ and $\{\Delta s_n\}$ are obtained. Subsequently, we consider the following dynamical variable [1,2]:

$$\eta_s(n) := \frac{\Delta U_{s_n}}{\sqrt{\Delta s_n}}, \tag{6}$$

which we referred to as the Loenwer driving force. From the analyses using the transfer operator method, for the arbitrary interface $\gamma_{[0,s]}$, the corresponding Loewner driving force $\eta_s(n)$ is topologically mixing. (See, Appendix I.) Here, the invariant measure $\rho_0^*$ is same as initial measure $\rho_o$ that is determined by the distribution of $\gamma_{[0,s]}$. In the physical sense, it indicates the stationarity and equilibrium of the Loewner driving force $\eta_s(n)$. Considering these facts, we define the Loewner entropy as the Boltzmann-type entropy of the Loewner driving force.

$$S_{\text{Loew}} := -\ln \rho(\eta_s(n)). \tag{7}$$

From the mixing property of $\eta_s(n)$, the Loewner entropy $S_{\text{Loew}}$ is additive, and the ensembles of the configurations of $\eta_s(n)$ are independent each other, even if the those of the Ising interface are correlated.

Subsequently, we consider the continuous limit of the 2D Ising interface at and below $T_c$. Since we observed that the discrete version of the interface $\gamma_{[0,s]}$ is described by discrete Loewner evolution whose driving force is expressed by Eq. (6), we reconstruct the driving function by taking the limit of the lattice segment size $\varepsilon \to 0$. Note that this limit corresponds to $\Delta s_n \to 0$ (See, Appendix I), and we consider the following Brownian motion-type process:

$$V_s = \lim_{\Delta s_n \to 0} \sum_{k=0}^{n} \eta_s(k). \tag{8}$$

The continuous Loewner evolution driven by $V_s$ defined above is described as follows:

$$\frac{\partial g_s(z)}{\partial s} = \frac{2}{g_s(z) - V_s}, \quad g_0(z) = z \in \mathbb{H}. \tag{9}$$

Although the reconstructed curve generated by Eq. (9) is different from the original Ising interface, the ensemble average of the Ising interface and the reconstructed curve becomes the same if the above-described continuous limit ($\varepsilon \to 0$ and $\Delta s_n \to 0$) is appropriately taken. From the mixing property of $\eta_s(n)$, which obeys a usual sense of central limit theorem (CLT), the behavior of the Loewner evolution in Eq. (9) is approximated by the SLE in the long-time limit; however, the geometry of the curve generated by Eq. (9) on $\mathbb{H}$ is quite different from that of the SLE because of the remaining relaxation process of the autocorrelation $\langle \eta_s(0)\eta_s(n) \rangle$. According to Ref. [3], the fractal dimension of the curve generated by the Loewner evolution in Eq. (9) is expressed as

$$d_f = 1 + K_{\sigma^2} \left[ \langle \eta_s(0)^2 \rangle + 2 \sum_{n=1}^{L} \langle \eta_s(0)\eta_s(n) \rangle \right], \tag{10}$$

where, $K_{\sigma^2}$ is a constant depending on the variance of $V_s$ expressed as $8/K_{\sigma^2} = \lim_{n \to \infty} \sigma^2{}_n$ where $\sigma^2{}_n$ is the variance of $(1/\sqrt{n}) \sum_{k=0}^{n} \eta_s(k)$ (See, Appendix II). Because the autocorrelation $\langle \eta_s(0)\eta_s(n) \rangle$ tends to vanish for the mixing dynamical systems for $n \to \infty$ [13], the fractal dimension in Eq. (10) becomes the same as that of the standard SLE, i.e., $d_f = 1 + (\kappa/8)$ [6] in the limit of $L \to \infty$; however, for the interface below $T_c$, there is the remaining autocorrelation for the range of summation in Eq. (10). For this fact, the fractal dimension of the interface at the critical temperature is different from that of the off-critical state for the finite lattice size $N \times N$. It should be noted that the fractal dimension is expressed as $d_f = 1 + K_{\sigma^2} \langle \eta_s(0)^2 \rangle$ at the critical temperature $T_c$, where the $V_s$ behaves as the Wiener process [3].

## B. Hamiltonian and grand canonical ensemble

In this subsection, we shall derive the Hamiltonian and grand canonical ensemble of the 2D Ising system, using the Loewner entropy and fractal dimension described above. The following theorem is one of the main results of this study.

**Theorem 1. (Hamiltonian)** *The energy (Hamiltonian) of the 2D Ising system at and below critical temperature $T_c$ is determined by Loewner entropy of the 2D Ising interface as*

$$H_N = \exp\left(\left\{1 + K_{\sigma^2}\left[\langle \eta_s(0)^2 \rangle + 2 \sum_{n=1}^{L} \langle \eta_s(0)\eta_s(n) \rangle\right]\right\} S_{\text{Loew}}\right) + C_N. \quad (11)$$

*where $K_{\sigma^2}$ is a constant parameter.*

*Proof.* The fractal dimension, or more precisely, the box-counting dimension of the 2D Ising interface is expressed as [3]:

$$d_f = \lim_{N,M \to \infty} \frac{\ln \text{len}(\gamma_{[0,s]}, N, M)}{\ln NM}. \quad (12)$$

where $N$ and $M$ are number of the small segment to cover the curve $\gamma_{[0,s]}$ along the imaginary axis and real axis, respectively. Using Eq. (2), Eq. (12) is expressed as the following:

$$d_f = \lim_{N,M \to \infty} \frac{\ln(H_N - C_N)}{\ln NM}. \quad (13)$$

From Eqs. (12) and (13), the Hamiltonian of the system is described as the following:

$$H_N = \exp\left(\left\{1 + K_{\sigma^2}\left[\langle \eta_s(0)^2 \rangle + 2 \sum_{n=1}^{L} \langle \eta_s(0)\eta_s(n) \rangle\right]\right\} \ln NM\right) + C_N. \quad (14)$$

Because in the limit of $\varepsilon \to 0$, $1/NM \simeq \rho(\varepsilon)\rho(\Delta U_\varepsilon) = \rho(\eta_s(n))$, Eq. (14) is rewritten as that including the term of Loewner entropy as Eq. (11).

In the above relation, the Hamiltonian is expressed as an exponential function of the autocorrelation of the Loewner driving force and Loewner entropy. The relation in Eq. (11) is an alternative expression to that described in Eqs. (1) and (2), and consistent with the numerical result reported in the previous literature [2]. Furthermore, the autocorrelation term can be modified using the Fourier transform $\eta_s(\omega) = \int_{-\infty}^{\infty} \eta_s(n) e^{-2\pi i \omega n} dn$. We consider the power spectral density of the Loewner driving force as:

$$E_{\text{Loew}}(\omega) = \lim_{t \to \infty} \frac{2\pi |\eta_s(\omega)|^2}{t}. \quad (15)$$

The similar quantity is studied in the context of mathematics as the *Loewner energy* [14].

**Remark 1.** *The relation in Eq. (11) is described in terms of the energy of the Loewner driving force*

as

$$H_N = \exp\left(\left\{1 + K_{\sigma^2}\left[\langle \eta_s(0)^2\rangle + 2\sum_{n=1}^{L}\int_{-\infty}^{\infty} E_{\text{Loew}}(\omega)e^{2\pi i\omega n}d\omega\right]\right\}S_{\text{Loew}}\right) + C_N, \quad (16)$$

where $E_{Loew}(\omega)$ is the power spectral density of the Loewner driving force expressed by Eq. (15).

*Proof.* Using the Wierner-Khinchin's theorem [15,16], autocorrelation term $\langle \eta_s(0)\eta_s(n)\rangle$ in Eq. (11) is rewritten as the following form:

$$\langle \eta_s(0)\eta_s(n)\rangle = \int_{-\infty}^{\infty} E_{\text{Loew}} e^{2\pi i\omega n} d\omega. \quad (17)$$

Substituting Eq. (17) into (14), the relation in Eq. (16) is obtained. It should be noted that for systems at the critical temperature $T_c$ the Loewner driving function $V_s$ corresponds to the Wiener process. It means that Loewner driving force $\eta_s(n)$ has no time correlation like the white noise. Therefore, we remark the following.

**Remark 2.** *At the critical temperature $T_c$, the autocorrelation term $\langle \eta_s(0)\eta_s(n)\rangle$ in Eq. (14) vanishes and the Hamiltonian is expressed as:*

$$H_N = \exp(\{1 + K_{\sigma^2}\langle \eta_s(0)^2\rangle\}S_{\text{Loew}}) + C_N. \quad (18)$$

*or*

$$H_N = \exp\left(\frac{11}{8}S_{\text{Loew}}\right) + C_N, \quad (19)$$

*for the continuous limit. The relation in Eq. (19) holds if the fractal dimension of the interface at $T_c$ is described by standard SLE with $\kappa = 3$ [10].*

Similarly, we argue that the grand canonical ensemble of the 2D Ising model is described in terms of the fractal dimension and Loewner entropy as the followings.

**Theorem 2. (Grand canonical ensemble):** *The grand canonical ensemble is expressed as the fractal dimension and Loewner entropy as:*

$$P(\gamma_{[0,s]}, \lambda_1, \lambda_2, \ldots, \lambda_n) = \frac{\exp\left[-\beta\left(\exp(d_f S_{\text{Loew}})\right)\right]}{Z(\Omega, \beta)} \quad (20)$$

*where,*

$$Z(\Omega, \beta) = \sum_{\gamma_s, \lambda_1, \lambda_2, \ldots, \lambda_n} \exp\left[-\beta\left(\exp(d_f S_{\text{Loew}})\right)\right]. \quad (21)$$

*Proof.* It follows from Eqs. (3), (4), and (11).

Furthermore, I remark that the configuration probability described by Eq. (20) is expressed by that of the curve for the subsequent discussions.

**Remark 3.** *On the assumption that the geometry of the interface and spin configuration have one-to-*

*one correspondence relation,*

$$P(\gamma_{[0,s]}, \lambda_1, \lambda_2, \ldots, \lambda_n) = P(\gamma_{[0,s]}) = P(z_0)P(z_0|z_1)\cdots P(z_{L-1}|z_L). \quad (22)$$

Here, we defined $P(\gamma_{[0,s]})$ as the path probability of the curve $\gamma_{[0,s]}$ which is expressed as the product of the conditional probability of the points on the curve. In the latter discussions, we denote $P(\gamma_{[0,s]})$ as $P_{path}(\gamma_{[0,s]})$.

### C. CLT for interface fluctuation

The important property of the local fluctuation of the 2D Ising interface has been investigated with its relation to the CLT [2, 8], which plays a key role to analyze the 2D Ising model. In our scheme, the variance $\sigma_{\gamma_s}^2$ of the local variation of the Ising interface on the real line is expressed as [2, 8]:

$$\sigma_{\gamma_s}^2 = \langle \Delta \mathrm{Re} z_0^2 \rangle + 2 \sum_{n \neq 0} \langle \Delta \mathrm{Re} z_0 \Delta \mathrm{Re} z_n \rangle. \quad (23)$$

Assuming the stationarity of $\eta_s(n)$, the probability distribution of the dynamics of the tip of the curve $z_n$ is the same as that of $\tilde{z}_s$ described by the Langevin equation expressed as [2]:

$$\frac{d\tilde{z}_s}{ds} = -\frac{2}{\tilde{z}_s} - \eta_s(n). \quad (24)$$

Using Eq. (24), in the long-time limit, the autocorrelation function of the local fluctuation on real axis is expressed as:

$$\langle \Delta \mathrm{Re} z_0 \Delta \mathrm{Re} z_n \rangle = \langle \Delta \mathrm{Re} \tilde{z}_0 \Delta \mathrm{Re} \tilde{z}_n \rangle = \langle \eta_s(0) \eta_s(n) \rangle. \quad (25)$$

Using the result on the Hamiltonian in Eq. (11), the following theorem is derived.

**Theorem 4. (CLT for interface):** *At and below the critical temperature $T_c$, the fluctuation of the interface of 2D Ising system is expressed by*

$$0 < \sigma_{\gamma_s}^2 = \frac{1}{K_{\sigma^2}}\left(\frac{\ln(H_N - C_N)}{S_{\mathrm{Loew}}} - 1\right) < \infty. \quad (26)$$

*Proof.* Eq. (23) can be rewritten in terms of the Loewner driving force as:

$$\sigma_{\gamma_s}^2 = \langle \eta_s(0)^2 \rangle + 2 \sum_{n \neq 0} \langle \eta_s(0) \eta_s(n) \rangle. \quad (27)$$

Substituting Eq. (27) into Eq. (11) we obtain:

$$H_N = \exp\{(1 + K_{\sigma^2} \sigma_{\gamma_s}^2) S_{\mathrm{Loew}}\} + C_N. \quad (28)$$

The following is immediately obtained from Eq. (28).

$$\sigma_{\gamma_s}^2 = \frac{1}{K_{\sigma^2}}\left(\frac{\ln(H_N - C_N)}{S_{\mathrm{Loew}}} - 1\right). \quad (29)$$

Furthermore, because of the mixing property of the Loewner driving force $\eta_s(n)$, wherein the CLT is valid, the variance $\sigma_{\gamma_s}^2$ is bounded as:

$$0 < \sigma_{\gamma_s}^2 < \infty. \tag{30}$$

From Eqs. (29) and (30), the relation in Eq. (26) is obtained.

In the followings, some important notions are remarked in relation to the above theorem.

**Remark 4.** *The condition $S_{Loew} = \ln(H_N - C_N)$ corresponds to the case that the interface is straight because it means $\sigma_{\gamma_s}^2 = 0$. This condition indicates the complete equilibrium.*

**Remark 5.** *The variance $\sigma_{\gamma_s}^2$ divergences if the autocorrelation $\langle \eta_s(0)\eta_s(n) \rangle$ does not vanishes for the long distance. It means that the ergodicity breaking in the Loewner driving force implies the phase transition.*

### D. Non-equilibrium free energy

The equilibrium state of the Ising system is closely related to the behavior of the free energy of the system. For the equilibrium condition, the free energy of the 2D Ising system is defined as $F_{eq} := -\beta^{-1} \ln Z(\Omega, \beta)$. In the followings, we consider the configuration probability as described by that of the curve $P_{path}(\gamma_{[0,s]})$ assuming that the contributions of the contour lines are invariant. We define the path probability of the curve $\gamma_{[0,s]}$ and driving function $V_s$ as [17]:

$$P_{path}(\gamma_{[0,s]}) = P(z_0)P(z_0|z_1)\cdots P(z_{L-1}|z_L) \tag{31}$$

and

$$P_{path}(V_s) = P(V_0)P(V_0|V_1)\cdots P(V_{L-1}|V_L). \tag{32}$$

Subsequently, the probability of the backward path of the curve $\gamma_{[0,s]}$ and driving function $V_s$ are defined as

$$\tilde{P}_{path}(\gamma_{[0,s]}) = P(z_L)P(z_L|z_{L-1})\cdots P(z_1|z_0) \tag{33}$$

and

$$\tilde{P}_{path}(V_s) = P(V_L)P(V_L|V_{L-1})\cdots P(V_1|V_0). \tag{34}$$

Here, we employ the central concept of stochastic thermodynamics described as the following [18]. We define the entropy production of the trajectory of the curve $\gamma_{[0,s]}$ and driving function $V_s$ as $\Delta S_{in}^\gamma$ and $\Delta S_{in}^V$, respectively as:

$$\Delta S_{in}^\gamma := \ln \frac{P_{path}(\gamma_{[0,s]})}{\tilde{P}_{path}(\gamma_{[0,s]})}, \qquad \Delta S_{in}^V := \ln \frac{P_{path}(V_s)}{\tilde{P}_{path}(V_s)}. \tag{35}$$

We note that the entropy production $\Delta S_{in}^\gamma$ and $\Delta S_{in}^V$ are assumed to be those inside the system, such that they are always non-negative from the second law of the thermodynamics. We define the non-equilibrium free energy of the 2D Ising system as the following:

$$F := E + \Delta E - TS(t), \tag{36}$$

where, $E$ and $S(t)$ are the internal energy and entropy, respectively. They are time dependent if the system is in non-equilibrium. $T$ denotes the constant temperature, and $\Delta E$ denotes the variation of the internal energy before it reaches the equilibrium.

**Theorem 4. (Non-equilibrium free energy):** *The non-equilibrium free energy of the 2D Ising system at and below the critical temperature $T_c$ defined in Eq. (36) is expressed as*

$$F = F_{eq} + \Delta E + k_B T \left\langle \frac{\Delta S_{in}^{\gamma} + \Delta S_{in}^{V}}{2} \right\rangle, \tag{37}$$

where, $\Delta E$ is the variation of the internal energy.

*Proof.* From the definition in Eq. (35), the following relation is derived:

$$\begin{aligned}\Delta S_{in}^{\gamma} + \Delta S_{in}^{V} &= \ln \frac{P_{path}(\gamma_{[0,s]})}{\tilde{P}_{path}(\gamma_{[0,s]})} + \ln \frac{P_{path}(V_{[0,s]})}{\tilde{P}_{path}(V_{[0,s]})} \\ &= \ln \frac{P_{path}(\gamma_{[0,s]})}{P_{path}(V_{[0,s]})} + \ln \frac{\tilde{P}_{path}(\gamma_{[0,s]})}{\tilde{P}_{path}(V_{[0,s]})}. \end{aligned} \tag{38}$$

Let us define the Kullback-Leibler (KL) divergence between the paths of the curve and driving function as:

$$D(\gamma_{[0,s]}|V_{[0,s]}) = \left\langle \ln \frac{P_{path}(\gamma_{[0,s]})}{P_{path}(V_{[0,s]})} \right\rangle, \quad \tilde{D}(\gamma_{[0,s]}|V_{[0,s]}) = \left\langle \ln \frac{\tilde{P}_{path}(\gamma_{[0,s]})}{\tilde{P}_{path}(V_{[0,s]})} \right\rangle. \tag{39}$$

Here, the ensemble average is taken over all of the possible interface (path) of the 2D Ising system. More explicitly, KL divergence in Eq. (39) is expressed as the following.

$$\begin{aligned} D(\gamma_{[0,s]}|V_{[0,s]}) &= \sum_{\gamma_{[0,s]} \in \Omega} P_{path}(\gamma_{[0,s]}) \ln P_{path}(\gamma_{[0,s]}) - \sum_{\gamma_{[0,s]} \in \Omega} P_{path}(\gamma_{[0,s]}) \ln P_{path}(V_{[0,s]}) \\ &= \sum_{\gamma_{[0,s]} \in \Omega} P_{path}(\gamma_{[0,s]}) \ln P_{path}(\gamma_{[0,s]}) - \sum_{V_{[0,s]} \in \Omega} P_{path}(V_{[0,s]}) \ln P_{path}(V_{[0,s]}) \\ &= -k_B^{-1}(S(t) + S_{eq}). \end{aligned} \tag{40}$$

From the one-to-one correspondence relation between the curve and driving function, we assumed that the ensemble average over $P_{path}(\gamma_{[0,s]})$ and $P_{path}(V_{[0,s]})$ is equivalent in Eq. (40). Further, the symmetric property of the curve and driving function leads to the following relation:

$$D(\gamma_{[0,s]}|V_{[0,s]}) = \tilde{D}(\gamma_{[0,s]}|V_{[0,s]}). \tag{41}$$

Using Eqs. (38), (39) and (41), the following is derived.

$$D(\gamma_{[0,s]}|V_{[0,s]}) = \frac{D(\gamma_{[0,s]}|V_{[0,s]}) + \tilde{D}(\gamma_{[0,s]}|V_s)}{2} = \left\langle \frac{\Delta S_{in}^{\gamma} + \Delta S_{in}^{V}}{2} \right\rangle. \tag{42}$$

From Eqs. (36), (40) and (42), we obtain Eq. (37).

**Remark 6.** *From the H-theorem* $\Delta F \leq 0$, *the variation of the energy* $\Delta E$ *is bounded as the following:*

$$\Delta E \leq -k_\beta T \left\langle \frac{\Delta S_{\text{in}}^\gamma + \Delta S_{\text{in}}^V}{2} \right\rangle. \tag{43}$$

The above relation determines the upper boundary of the variation of the internal energy of the non-equilibrium state of 2D Ising system.

### E. Fluctuation-dissipation relation

Consider the fluctuation-dissipation relation (FDR) for the 2D Ising system. If the system subjected to the small perturbation, the response in the system should be theoretically predicted. The usual FDR is based on the linear response formalism using the correlation function. For the non-equilibrium Ising systems, some studies have revealed the modified FDR [19]. In this study, we assume that the perturbation to the Ising system is locally to the near around the interface. We estimate the effect of the perturbation to the interface to the interface at the distant point. Let us define the response function:

$$R(s, s') = \frac{\langle dz_s \rangle}{dc_{s'}}, \qquad s > s', \tag{44}$$

where, $dc_{s'}$ is the perturbation at the point $s'$ of the interface, and $\langle dz_s \rangle$ is the variation of the interface after the perturbation occurs. By applying the linear response formalism with respect to the Loewner driving force, we obtained the following theorem.

**Theorem 5. (FDR):** *The local perturbation to near the interface curve* $\gamma_{[0,s]}$ *results in the nonlinear variation in the 2D Ising interface, whose response function is described as:*

$$R(s, s') = -\beta \sum_{\gamma_s \in \Omega} V_s(n) \eta_s(n') \exp(-S_{\text{Loew}}), \tag{45}$$

*if* $V_s(n)$ *has a stationary distribution.*

*Proof.* From Eq. (24) the response function $R(s, s')$ corresponds to that of $V_s(n)$ for sufficiently large $s$. Therefore, the response function in Eq. (44) is rewritten as that of $V_s(n)$. Assuming the stationary property of $V_s(n)$, $R(s, s')$ is expressed as the linear response function described as [19-21]:

$$R(s, s') = -\beta \langle V_s(n) \Delta V_s(n') \rangle_{\text{eq}}. \tag{46}$$

Here, the ensemble average is taken over $P(V_s(n))$. Using the fact that both of $V_s(n)$ and $\eta_s(n)$ have the stationary distribution obeying the CLT, Eq. (46) is rewritten as:

$$R(s, s') = -\beta \sum_{\gamma_s \in \Omega} P(\eta_s(n)) V_s(n) \eta_s(n'). \tag{47}$$

Using the definition of Loewner entropy in Eq. (7), we obtain $P(\eta_s(n)) = \exp(-S_{\text{Loew}})$, and substituting this into Eq. (47), the relation in Eq. (45) is derived.

**Remark 7.** *At the critical temperature $T_c$, the response function $R(s, s')$ diverges because $V_s(n)$ is not stationary. In this case, the prediction of the nonlinear response using Eq. (45) is difficult.*

The above statement means that Theorem 5 retains its validity for off-critical state of the 2D Ising model ($T < T_c$). Although we here described the response as the coordinate change in the point of the interface, it is caused by the change in the spin configuration near the reference point $z_s$ and $z_{s'}$ of the interface.

## IV. CONCLUSION

In this article, the statistical mechanics of the 2D Ising model was reformulated using the Loewner entropy, which is defined as the entropy of the Loewner driving forces corresponding to the interface of the 2D Ising model. We observed some relations based on the assumption that the discrete Loewner evolution is regarded as a dynamical system having an invariant probability measure, which was demonstrated using the transfer operator method [22]. Using this result, the Loewner evolution driven by the random walk composed from mixing (chaotic) sequence was reconstructed and the fractal dimension of the curve was derived to compare it with the Hamiltonian, grand canonical ensemble, and local fluctuation of the interface. It was found that these are expressed in terms of Loewner entropy and fractal dimension of the interface in the continuous limit of the lattice segment size $\varepsilon \to 0$ and the increment of the Loewner time $\Delta s_n \to 0$. It should be remarked that because $\{\Delta s_n\}$ in the discrete Loewner evolution is heterogeneous sequence, this continuous limit is one of the possible forms, which might be different from the usual sense of it.

We further discussed the CLT for interface fluctuation in terms of the mixing property of the Loewner driving force, whose variance plays an important role for the physical properties of the 2D Ising system such as complete equilibrium state and phase transition. The obtained results indicate that the off-critical Ising interface also is described in terms of the usual sense of CLT if we assume the convergence of the autocorrelation function of the Loewner driving force for $n \to \infty$. In addition, the non-equilibrium free energy for the 2D Ising system was derived using the approach of the stochastic thermodynamics. The derived relation was based on the path probability formalism and determines the boundary of the energy variation in the 2D Ising system with Dobrushin boundary condition. Further, we derived the FDR for the 2D Ising system using Loewner entropy. For deriving this relation, we used the stationarity of the driving function, which is possible when the diffusivity speed of the driving function is less than Wiener process ($\leq n^1$) due to the autocorrelation in the Loenwer driving force. We remark that the formulation using Loewner theory presented in this study can also be applied to one-dimensional dynamics in non-equilibrium state if we regard their orbit as curve morphology. (Such an attempt is performed in for example in Ref. [23].) Particularly, the "complexity-energy" relationship we demonstrated using Loewner entropy, is worth investigating in more detail dealing with the concrete models in non-equilibrium states. Therefore, the applications of the present approach to various non-equilibrium systems are required for the unified understanding of the statistical mechanical models, which will be performed in the future studies.

APPENDIX I. Mixing Property of Discrete Loewner Evolution

The mixing property of the discrete Loewner evolution has been shown in Ref. [22] for the theoretical investigation of the Loewner time conversion to the anomalous diffusion. We here demonstrate that for an even more general case, i.e., the tip of the curve moves along the imaginary axis freely, the topological mixing property holds true. We denote the coordinates of the interface $\gamma_{[0,s]} = \{z_0 = 0, z_1, z_2, \dots, z_n, \dots, z_L\}$ as

$$z_n = \sum_{k=0}^{n} \left(\sqrt{-1}\right)^m \varepsilon(k), \qquad m \in \{1, 2, 3, 4\}. \tag{A1}$$

with $z_0 = 0$, $z_L = (0, N/2)$, and $\varepsilon(k)$ is a lattice segment size, which is constant and not dependent on the index $k$. The choral Loewner evolution described in the above section is based on the continuous time differential equation. For the practical applications of the (continuous) Loewner evolution requires an appropriate discretization of Loewner evolution [11]. The most simple and well-used method is the zipper-algorithm based on the vertical slit map [12]. This method is implemented by the conformal map called 'vertical slit map', which maps the small vertical segment on $\mathbb{H}$ to a point on the real axis. Let us consider the map:

$$g_s(z) = \varepsilon + \sqrt{(z - \varepsilon)^2 + 4s} \in \mathbb{H}. \tag{A2}$$

This map transforms the region $\mathbb{H}$ minus the small vertical segment having the length $\varepsilon$ to $\mathbb{H}$ and the tip of the segment is moved to $\varepsilon$ on the real axis. This map is a solution of the Loewner equation in Eq. (5), and an element for obtaining the discrete version of the chordal Loewner evolution.

For discretizing the curve $\gamma_{[0,s]}$ and the driving function $U_s$, the time interval $0 < s_1 < s_2 <$

$\cdots < s_n < \cdots < s_{L-1} < s_L$. Accordingly, the driving function $U_s$ is indexed by this time interval as: $0 < U_{s_1} < U_{s_2} < \cdots < U_{s_n} < \cdots U_{s_{L-1}} < U_{s_L}$. We denote the increment of $s_n$ as $\Delta s_n := s_n - s_{n-1}$ and that of $U_{s_n}$ as $\Delta U_{s_n} := U_{s_n} - U_{s_{n-1}}$.

$$g_n(z) = \Delta U_{s_n} + \sqrt{(z - \Delta U_{s_n})^2 + 4\Delta s_n}. \tag{A3}$$

Subsequently, let us define the shifted map of $g_n(z)$ as the following:

$$h_n(z) := g_n(z) - \Delta U_{s_n} = \sqrt{(z - \Delta U_{s_n})^2 + 4\Delta s_n}. \tag{A4}$$

Considering the curve $\gamma_{[0,s]}$ is expressed by the discretized point on $\mathbb{H}$ as $\gamma_{[0,s]} = \{z_0 (= 0), z_1, \ldots, z_n, \ldots, z_L\}$, from the composition property of Loewner evolution discussed in Ref. [5], the following relation is obtained.

$$h_n \circ h_{n-1} \circ \cdots h_1(z_n) = 0. \tag{A5}$$

If we stop the iteration of $h_n$ to $z_n$ one step before it reaches to zero, a small vertical segment remains the upper half-plane. The Kennedy's algorithm [12] focuses on the fact that whose tip is located on the following variable, which consists of the increments of Loewner driving function:

$$w_n = \Delta U_{s_n} + 2i\sqrt{\Delta s_n}. \tag{A6}$$

Here, we note that $i = \sqrt{-1}$. The algorithm to obtain the discrete Loewner driving process from the points on the curve is expressed as:

$$w_1 = z_1$$
$$w_2 = h_1(z_2)$$
$$\vdots$$
$$w_n = h_{n-1} \circ h_{n-2} \circ \cdots \circ h_1(z_n)$$
$$\vdots$$
$$w_L = h_{L-1} \circ h_{L-2} \cdots \circ h_n \cdots \circ h_1(z_L). \tag{A7}$$

In this study, we regard the above transformation as *n*-dependent dynamical system, whose initial condition is given by the position of the curve $\gamma_{[0,s]}$. In the next section, using the transfer operator method, we analyze the dynamical system property of the above transformation.

Let us define the transfer operator $\mathcal{L}_n$ as that satisfying the following [24]:

$$\mathcal{L}_n \rho(w) = \sum_{z \in h^{-1}(w)} |h_n'(\chi_\sigma(w))|^{-1} \rho(\chi_\sigma(w)). \tag{A8}$$

Here, the sum is taken over all of the preimages $\chi_\sigma(w)$ of $h_n(z)$ expressed as:

$$\chi_\sigma(w) := h^{-1}_{\sigma=1,2} = \Delta U_{s_n} \pm \sqrt{w^2 - 4\Delta s_n}, \tag{A9}$$

and we obtain

$$|h_n'(\chi_\sigma(w))|^{-1} = \left|\frac{\pm\sqrt{w^2 - 4\Delta s_n}}{w}\right|^{-1}. \tag{A10}$$

For a fixed $n$, the action of the $\mathcal{L}_n$ to the probability density function of $w_n$, denoted as $\rho(w)$ is expressed as the following:

$$\rho_N(w) = \mathcal{L}_n{}^L \rho_0(w) \tag{A11}$$

Here, $\rho_0(w)$ denotes the initial distribution and $\rho_N(w)$ denote that obtained after $L$ times iterations. With the above settings, we derive the following theorem.

**Theorem A1. (Topological Mixing):** *The transformation in Eq. (3.6) to the interface $\gamma_{[0,s]}$ of the 2D Ising model is topologically mixing, in the sense that,*

$$|\rho_L(w) - \rho_0^*| \sim \exp[-K(L)], \quad \text{and} \quad K(L) \to \infty \text{ as } L \to \infty. \tag{A12}$$

*where $\rho_0^*$ is the invariant probability density and $K(L)$ is a function of $L$.*

*Proof.* Using Eqs. (A8) and (A10), Eq. (A11) is expressed as :

$$\rho_L(w) = \sum_{\sigma=1,2} \rho_0(\chi_\sigma(w)) |h_n{}^{L'}(\chi_\sigma(w))|^{-1}$$

$$= \rho_0(\chi_\sigma(w)) \sum_{\sigma=1,2} \exp\left[-\sum_{n=0}^{L-1} \ln|h_n'(\chi_\sigma(w))|\right]$$

$$= \rho_0(\chi_\sigma(w)) \sum_{\sigma=1,2} \exp\left[-\sum_{n=0}^{L-1} \ln\left|\frac{\pm\sqrt{w^2 - 4\Delta s_n}}{w}\right|\right]. \tag{A13}$$

Denoting $K(L) := \sum_{n=0}^{L-1} \ln\left|\frac{\pm\sqrt{w^2 - 4\Delta s_n}}{w}\right|$, we shall demonstrate that $K(L) \to \infty$ for $L \to \infty$. From Eqs. (A1) and (A7), it is observed that

$$h_1(z_n) = \sqrt{\left[\text{Re}\left(\sum_{k=0}^{n}(\sqrt{-1})^m \varepsilon(k)\right) + i\text{Im}\left(\sum_{k=0}^{n}(\sqrt{-1})^m \varepsilon(k)\right) - \Delta U_{s_1}\right]^2 + 4\Delta s_1} \tag{A14}$$

For the subsequent calculation, we denote

$$a_1 = \text{Re}\left[\text{Re}\left(\sum_{k=0}^{n}(\sqrt{-1})^m \varepsilon(k)\right) + i\text{Im}\left(\sum_{k=0}^{n}(\sqrt{-1})^m \varepsilon(k)\right) - \Delta U_{s_1}\right]^2$$

$$= \text{Re}\left(\sum_{k=0}^{n}(\sqrt{-1})^m \varepsilon(k)\right)^2 + \text{Im}\left(\sum_{k=0}^{n}(\sqrt{-1})^m \varepsilon(k)\right)^2 + \Delta U_{s_1}^2$$

$$- 2\text{Re}\left(\sum_{k=0}^{n}(\sqrt{-1})^m \varepsilon(k)\right)\Delta U_{s_1}. \tag{A15}$$

and

$$a_2 = \text{Im}\left[\text{Re}\left(\sum_{k=0}^{n}(\sqrt{-1})^m \varepsilon(k)\right) + i\text{Im}\left(\sum_{k=0}^{n}(\sqrt{-1})^m \varepsilon(k)\right) - \Delta U_{s_1}\right]^2$$

$$= 2\left[\text{Re}\left(\sum_{k=0}^{n}(\sqrt{-1})^m \varepsilon(k)\right)\text{Im}\left(\sum_{k=0}^{n}(\sqrt{-1})^m \varepsilon(k)\right) + \text{Im}\left(\sum_{k=0}^{n}(\sqrt{-1})^m \varepsilon(k)\right)\Delta U_{s_1}\right]. \tag{A16}$$

Using Eqs. (A15) and (A16), Eq. (A14) is rewritten as:

$$h_1(z_n) = \sqrt{a_1 + ia_2}$$

$$= \pm\sqrt{\frac{a_1 \pm \sqrt{a_1^2 - a_2^2}}{2}} \mp i\sqrt{2}\left(\frac{a_1\sqrt{a_1 - \sqrt{a_1^2 - a_2^2}} \pm \sqrt{a_1^2 + a_2^2}\sqrt{a_1 \pm \sqrt{a_1^2 - a_2^2}}}{2a_2}\right)$$

$$= \pm\sqrt{\frac{a_1 \pm \sqrt{a_1^2 - a_2^2}}{2}} + iO(\sqrt{n\varepsilon}). \tag{A17}$$

Here, $i = \sqrt{-1}$. Applying the recursive algorithm in Eq. (A7), we obtain,

$$h_2 \circ h_1(z_n) = h_2\left(\pm\sqrt{\frac{a_1 \pm \sqrt{a_1^2 - a_2^2}}{2}} + iO(\sqrt{n\varepsilon})\right)$$

$$= \pm\sqrt{\frac{a_3 \pm \sqrt{a_3^2 - a_4^2}}{2}} + iO(\sqrt{n\varepsilon}). \tag{A18}$$

Repeating the above procedure, the variable $w_n$ is expressed by the following:

$$w_n = h_{n-1} \circ h_{n-2} \cdots \circ h_n \cdots \circ h_1(z_n) = f(a_1, a_2, \ldots, a_{2n}) + iO(\sqrt{n\varepsilon}), \tag{A19}$$

where $f(a_1, a_2, \ldots, a_{2n})$ is a real-valued function. From Eqs. (A6) and (A19), we observe that

$$\Delta s_n \to \infty \quad \text{as} \quad \sqrt{n\varepsilon} \to \infty. \tag{A20}$$

Thus, from Eqs (A10) and (A20), we obtain the following

$$K(L) = \sum_{n=0}^{L-1} \ln\left|\frac{\pm\sqrt{w^2 - 4\Delta s_n}}{w}\right| \to \infty \quad \text{as} \quad L \to \infty. \tag{A21}$$

From Eqs. (A13) and (A21), the initial distribution $\rho_0$ corresponds to the invariant density $\rho_0^*$, and

Eq. (A12) is obtained.

This result indicates that the transformation in Eq. (A7) is topologically mixing. Subsequently, we define the Loewner driving force as:

$$\eta_s(n) := \frac{\Delta U_{s_n}}{\sqrt{\Delta s_n}}. \quad (A22)$$

For the Theorem A1, the random sequence $\{\eta_s(n)\}$ also is topologically mixing. If we take the continuous limit $\varepsilon \to 0$, the random walk-like driving function is defined as:

$$V_s = \lim_{\Delta s_n \to 0} \sum_{k=0}^{n} \eta_s(k). \quad (A23)$$

The property of the Loenwer driving force and driving function was numerically investigated in Refs. [1-3].

APPENDIX II. Fractal Dimension

The fractal dimension of the chaotic analog of SLE is derived in Ref. [3], based on the method described in Ref. [25]. In the followings we describe the method to derive the fractal dimension of the curve generated by Loewner evolution in Eq. (9) that is same as those in the previous literature. We note that the important point is the convergence of the autocorrelation of the Loewner driving force $\eta_s(n)$. If we consider the system of the finite size $N$, wherein the $\langle \eta_s(0)\eta_s(n)\rangle$ does not converge completely, the following argument remains to be valid because of the symmetric property of the Loewner driving function $V_s$.

Let $\rho(x, y, \varepsilon)$ be the probability density function expressing the probability that the point $(x, y)$ exits within the distance $\varepsilon$ from the curve $\gamma_{[0,s]}$. Consider the conformal map corresponding to the infinitesimal evolution as $g_\varepsilon(z)$. Using the relations, $dx = \frac{2x}{x^2+y^2}dt, dy = -\frac{2y}{x^2+y^2}dt$, $|g_\varepsilon'(z)| \sim [1 - 2dt\,\mathrm{Re}\left(\frac{1}{z^2}\right)]\varepsilon$, and $\langle dx^2 \rangle = \sqrt{K_{\sigma^2}/8}\langle [\eta_s(n)]^2 \rangle$, we observe that, under this map, the probability density function $\rho(x, y, \varepsilon)$ changes as follows:

$\rho(x, y, \varepsilon) = \rho(x + dx, y + dy, \varepsilon + d\varepsilon)$

$$\simeq \rho(x,y,\varepsilon) + \left[\frac{K_{\sigma^2}\sigma^2{}_n}{16}\frac{\partial^2}{\partial x^2} + \frac{2x}{x^2+y^2}\frac{\partial}{\partial x} - \frac{2y}{x^2+y^2}\frac{\partial}{\partial y} - 2\mathrm{Re}\left(\frac{1}{z^2}\right)\varepsilon\frac{\partial}{\partial \varepsilon}\right]\rho(x,y,\varepsilon). \quad (A24)$$

Here $\sigma^2{}_n$ is the variance of $(1/\sqrt{n})\sum_{k=0}^{n}\eta_s(k)$. From the symmetric property of the driving function, we assume that the measure describing the curve does not change under the map $g_\varepsilon(z)$.

Using this, Eq. (A24) leads to the following:

$$\left[\frac{K_{\sigma^2}\sigma^2{}_n}{16}\frac{\partial^2}{\partial x^2} + \frac{2x}{x^2+y^2}\frac{\partial}{\partial x} - \frac{2y}{x^2+y^2}\frac{\partial}{\partial y} - 2\text{Re}\left(\frac{1}{z^2}\right)\varepsilon\frac{\partial}{\partial \varepsilon}\right]\rho(x,y,\varepsilon) = 0. \quad (A25)$$

According to Ref. [25], the ansatz solving the above equation is described as following:

$$\rho(x,y,\varepsilon) = \varepsilon^{2-d_f} x^\alpha (x^2+y^2)^\beta. \quad (A26)$$

To regard $\rho(x,y,\varepsilon)$ as dimensionless function, we obtain the following relationship.

$$\alpha + 2\beta = d_f - 2. \quad (A27)$$

From Eqs. (A25) and (A26), we obtain,

$$\alpha = \frac{[(K_{\sigma^2}\sigma^2{}_n/8) - 8]^2}{8\sigma^2{}_n}, \quad \beta = \frac{(K_{\sigma^2}\sigma^2{}_n/8) - 8}{2\sigma^2{}_n}. \quad (A28)$$

From Eqs. (A27) and (A28), the fractal dimension $d_f$ is described as

$$d_f = 1 + K_{\sigma^2}\sigma^2{}_n. \quad (A29)$$

Here, $K_{\sigma^2}$ works as a normalization factor of the variance of the chaotic sequence $\langle[\eta_s(n)]^2\rangle$. From the central limit theorem (CLT), $\sigma^2{}_n$ is expressed as [13]:

$$\sigma^2{}_n = \langle\eta_s(0)^2\rangle + 2\sum_{n=1}^{L}\langle\eta_s(0)\eta_s(n)\rangle. \quad (A30)$$

Using Eqs. (A29) and (A30), we obtain Eq. (10).

ACKNOWLEDGEMENTS

This work was performed during my stay in IHES. I wish to express my sincere gratitude for their stimulus environment and hospitality.